\documentclass[12pt]{iopart}

\usepackage{amssymb}
\usepackage{iopams}
\usepackage{algorithm}
\usepackage{algorithmic}
\begin{document}

\title[The Origin of Separable States and Separability Criteria from Entanglement-breaking Channels]{The Origin of Separable States and Separability Criteria from Entanglement-breaking Channels}
\author{Banghai Wang$^{1,2}$, Qin Li$^1$ and Dongyang Long$^1$}
\address{$^1$Department of
Computer Science, Sun Yat-sen University, Guangzhou 510006,
P.R.China}
\address{$^2$Faculty of Computer, Guangdong University of
technology, Guangzhou 510006, P.R.China} \ead{wangbanghai@gmail.com,
liqin805@163.com, issldy@mail.sysu.edu.cn}
\begin{abstract}
In this paper, we show that an arbitrary separable state can be the
output of a certain entanglement-breaking channel corresponding
exactly to the input of a maximally entangled state. A necessary and
sufficient separability criterion and some sufficient separability
criteria from entanglement-breaking channels are given.

\end{abstract}

\maketitle

\section{Introduction}
Quantum entanglement, which is applied to various types of quantum
information processing such as quantum computation\cite{Shor94},
quantum dense coding\cite{Bennett92}, quantum
teleportation\cite{Bennett93}, quantum cryptography\cite{Ekert91},
etc., lies in the heart of quantum information theory. However,
quantum entanglement is not yet fully understood by people, and many
problems on entanglement remain open\cite{Horodecki09}. It is one of
the central problems to check whether or not a state is entangled.
Some
necessary\cite{Peres96,R.Horodecki96,Horodecki97,Horodecki99,Nielsen01,Rudolph03,Chen03,Guhne07}
or sufficient\cite{Braunstein99,Pittenger00}separability criteria,
as well as the necessary and sufficient separability criterion for
low dimension states\cite{M.Horodecki96}, have been found.
Unfortunately, there has not been any effective necessary and
sufficient separable condition for an arbitrary state yet.

Entanglement underlines the intrinsic order of statistical relations
between subsystems of a composite quantum
system\cite{Horodecki09,Schrodinger35}. In this paper, we show that
there exists a correlation between the separable state and a
maximally entangled state in any quantum systems of arbitrary
dimensions from the entanglement-breaking channel (EBC). This paper
is organized as follows. Sec. II demonstrates that all separable
states can be the output of certain EBCs corresponding exactly to
the input of a maximally entangled state. The separability criteria
from EBCs are investigated in Sec. III. Sec. IV summarizes our main
results.

\section{The origin of all separable states can be a maximally entangled state}

Let $\mathcal{H}{_{A}}, \mathcal{H}{_{B}}$ be two Hilbert spaces,
$\mathcal{H}_{A,B}$ be the tensor product space
$\mathcal{H}{_{A}}\otimes\mathcal{H}{_{B}}$. Denote
$\mathcal{B}(\mathcal{H}_{A,B})$ as the set of operators on
$\mathcal{H}_{A,B}$, $\mathcal{B}(\mathcal{H}_{A,B})^{+}$ as the
subset of positive semidefinite operators (i.e. the unnormalized
density operators) on $\mathcal{H}_{A,B}$. Let
$\dim\mathcal{H}_{A}={d}_{A}$ and $\dim\mathcal{H}_{B}={d}_{B}$. A
mixed state $\rho\in \mathcal{B}(\mathcal{H}_{A,B})^{+}$ is called
separable state if it can be written as
 \begin{eqnarray}
\rho=\sum_{k} p_{k}|\psi_{k}\rangle\langle\psi_{k}|\otimes
 |\phi_{k}\rangle\langle\phi_{k}|,
\end{eqnarray}
where $\{p_k\}$ is a probability distribution, $|\psi_{k}\rangle$
and $|\phi_{k}\rangle$ are pure states of $\mathcal{H}_{A}$ and
$\mathcal{H}_{B}$ respectively.

A quantum channel is a completely positive (CP) linear
map\cite{Kraus83,Verstraete03}. The EBC is the class of quantum
channels, for which $(I\otimes \Phi)(\rho)$ is always separable for
any state $\rho$, where $\rho$ is separable or
not\cite{Holevo99,M.Horodecki03,Ruskai03}. By Ref.\cite{Ruskai03},
an EBC can be written in either of the following equivalent forms
\begin{eqnarray}
\Phi(\sigma)&=&\sum_{k}R_{k}TrF_{k}\sigma \\
&=&\sum_{k}|\psi_{k}\rangle\langle\psi_{k}|\langle\phi_{k}|\sigma|\phi_{k}\rangle,
\end{eqnarray}
where $\sigma$ is any density operator, $R_{k}$ is a density
operator and $F_{k}$ is a positive semi-definite operator. The
channel $\Phi$ is entanglement-breaking and trace-preserving(EBT) if
and only if
$\sum_{k}F_{k}=\sum_{k}|\phi_{k}\rangle\langle\phi_{k}|=I$, where
the set $\{F_{k}\}$ form a POVM. The properties of EBT were
investigated by many
people\cite{Holevo99,M.Horodecki03,Ruskai03,Shor04}. The action of
an EBT can be substituted by a measurement and state-preparation
protocol\cite{Korbicz08}: The sender makes a measurement on the
input state $\sigma$ by means of a POVM $\{F_{k}\}$, and sends the
outcome $k$ via a classical channel to the receiver. Then the
receiver prepares an agreed-upon state
$R_{k}$\cite{M.Horodecki03,Sacchi05}.

However, a general channel need be not trace-preserving, and a
channel is trace-preserving if no loss of the
particle\cite{Kraus83,Verstraete03}. Any finite-dimensional
completely positive trace-preserving (CPT) linear map $\Phi$ can be
represented as
\begin{eqnarray}
\Phi(\sigma)=\sum_{k}E_k\sigma E_k^\dagger,
\end{eqnarray}where the $E_k$ are complex matrices satisfying
\begin{eqnarray}
\sum_{k}E_kE_k^\dagger=I.
\end{eqnarray}
In this paper, we consider
\begin{eqnarray}
\sum_{k}E_kE_k^\dagger\leq I,
\end{eqnarray}i.e. the channel has only the completely positive
property and need not trace-preserving. Some properties hold for the
channel without trace-preserving but not for the channel with
trace-preserving, vice verse.

We need two easily proved lemmas.

\textbf{Lemma 1.}\cite{Li09} A pure bipartite state
$|\beta\rangle\in\mathbb{C}^{d_A}\otimes\mathbb{C}^{d_B}(d_{B}\geq
d_A)$ is a maximally entangled if and only if
$E(|\beta\rangle)=\log_2d_A$, where
$E(|\beta\rangle)=-\textrm{Tr}(\rho_{A(B)}\log_2\rho_{A(B)})$ and
$\rho_{A(B)}=\textrm{Tr}_{B(A)}(|\beta\rangle\langle\beta|)$ is the
reduced density operator.

\textbf{Lemma 2.}\cite{Li09} A pure bipartite state
$|\beta\rangle\in\mathbb{C}^{d_A}\otimes\mathbb{C}^{d_B} (d_{B}\geq
d_A)$ is a maximally entangled state if and only if there exits an
orthonormal basis in $\mathbb{C}^{d_B}$ for any given orthonormal
complete basis $\{|i_A\rangle\}$ in $\mathbb{C}^{d_A}$, such that
$|\beta\rangle$ can be written in the following form
\begin{eqnarray}
\label{puremaximallyentangled}
|\beta\rangle=\frac{1}{\sqrt{d_A}}\sum_{i=1}^{d_A}|i_A\rangle\otimes|i_B\rangle.
\end{eqnarray}

\textbf{Theorem 1.} An arbitrary separable state can be the output
of a certain EBC corresponding exactly to the input of a maximally
entangled state.

 \textbf{proof:} Without loss of generality,
 suppose an arbitrary separable state $\rho=\sum_{k} p_{k}|\psi_{k}\rangle\langle\psi_{k}|\otimes
 |\phi_{k}\rangle\langle\phi_{k}|$ in $\mathbb{C}^{d_A}\otimes \mathbb{C}^{d_B}$ and $d_B\geq
d_A$. Since $|\psi_{k}\rangle$
 can be represented as a linear combination of the orthomormal basis $\{|i_A\rangle\}$ in $\mathbb{C}^{d_A}$, i.e. $|\psi_{k}\rangle=\sum_{i}|i_A\rangle\langle i_A|\psi_{k}\rangle$, we have
 \begin{eqnarray}
\rho&=&\sum_{k} p_{k}(\sum_{i}|i_A\rangle \langle
i_A|\psi_{k}\rangle)(\sum_{j}\langle j_A |\langle
j_A|\psi_{k}\rangle)|\otimes
 |\phi_{k}\rangle\langle\phi_{k}| \\
&=&\sum_{i,j,k}|i_A\rangle\langle j_A|\otimes
|\phi_{k}\rangle\langle\phi_{k}|p_{k}\langle
i_A|\psi_{k}\rangle\langle j_A|\psi_{k}\rangle\\
&=&\sum_{i,j}|i_A\rangle\langle
j_A|\otimes\sum_{k}|\phi_{k}\rangle\langle\phi_{k}|\textrm{Tr}(|i_A\rangle\langle
j_A|p_{k}|\psi_{k}\rangle\langle\psi_{k}|)\\
&=&\sum_{i,j}|i_A\rangle\langle
j_A|\otimes\sum_{k}|\phi_{k}\rangle\langle\phi_{k}|\textrm{Tr}(|i_B\rangle\langle
j_B|p_{k}|\psi_{k}\rangle\langle\psi_{k}|)\\
&=&\sum_{i,j}|i_A\rangle\langle j_A|\otimes\Phi(|i_B\rangle\langle j_B|)\\
&=&(I\otimes\Phi)(|I\rangle\langle I|),
\end{eqnarray} where $\Phi(\sigma)=\sum_{k}|\phi_{k}\rangle\langle\phi_{k}|\textrm{Tr}(\sigma
F_{k}) (F_{k}=p_{k}|\psi_{k}\rangle\langle\psi_{k}|)$ is an EBC,
$|I\rangle=\sum_{i}|i_A\rangle\otimes|i_B\rangle$ is an unnormalized
maximally entangled state.\hfill $\square$

Now we give an interpretation of EBC with the input of a maximally
entangled state. Since
$\sum_{k}F_{k}=\sum_{k}p_{k}|\psi_{k}\rangle\langle\psi_{k}|\lneq
I$, $\Phi(\cdot)$ is non-trace-preserving and does not provide a
complete description of the processes that may occur in the
system\cite{Nielsen00}, the output separable state may be gotten
with some probability. Let the separable state, for example,
$\rho=|00\rangle\langle00|$. With the input of the two-qubit
maximally entangled state (Bell state)
$\frac{1}{2}(|00\rangle\langle00|+|00\rangle\langle11|+|11\rangle\langle00|+|11\rangle\langle11|)$,
we have
 \begin{eqnarray}
&&I\otimes\Phi(\frac{1}{2}(|00\rangle\langle00|+|00\rangle\langle11|+|11\rangle\langle00|+|11\rangle\langle11|))\nonumber\\
&=&|0\rangle\langle0|\otimes
(|0\rangle\langle0|\textrm{Tr}((\frac{1}{2}
(|0\rangle\langle0|+|0\rangle\langle1|+|1\rangle\langle0|+|1\rangle\langle1|))(|0\rangle\langle0|)))\nonumber\\
&=&\frac{1}{2}|0\rangle\langle0|\otimes|0\rangle\langle0|\nonumber\\
&=&\frac{1}{2}|00\rangle\langle00|,\nonumber
 \end{eqnarray}
i.e. the result $|00\rangle\langle00|$ by EBC occurs with the
probability of $\frac{1}{2}$.

Rewriting Eq. (11), we have
 \begin{eqnarray}
\rho&=&\frac{1}{d_{A}}\sum_{i,j}|i_A\rangle\langle j_A|\otimes
d_{A}\sum_{k}|\phi_{k}\rangle\langle\phi_{k}|\textrm{Tr}(|i_B\rangle\langle
j_B|p_{k}|\psi_{k}\rangle\langle\psi_{k}|)\\
&=&\frac{1}{d_{A}}\sum_{i,j}|i_A\rangle\langle j_A|\otimes\Phi^{'}(|i_B\rangle\langle j_B|)\\
&=&(I\otimes\Phi^{'})(|\beta\rangle\langle\beta|),
\end{eqnarray} where $\Phi^{'}(\sigma)=d_{A}\sum_{k}|\phi_{k}\rangle\langle\phi_{k}| \textrm{Tr}(\sigma
F^{'}_{k})$, $F^{'}_{k}=d_{A}p_{k}|\psi_{k}\rangle\langle\psi_{k}|$,
and
$|\beta\rangle=d_{A}^{-1/2}\sum_{i}|i_A\rangle\otimes|i_B\rangle$ is
the normalized maximally entangled state. Clearly, only if one of
reduced sate and all reduced sates of the separable state are
maximally mixed, $\Phi^{'}(\cdot)$ is EBT, otherwise
$\Phi^{'}(\cdot)$ is non-trace-preserving. The theoretical and
experimental analysis of non-trace-preserving processes have been
carried out\cite{Kiesel05,Kofman09,Bongioanni10}.

\section{The separability criteria from entanglement-breaking channels}
By the definition of EBC, we can easily get the separability
criteria from entanglement-breaking channels.

\textbf{Theorem 2.} A state is entangled if and only if it is not
any output of the EBC.

It is impossible to travel through all EBCs, and therefore the
separability criterion above is not operational. However, from the
definition of EBC, we can get a class of sufficient separability
criteria in matrix form.
\subsection{The separability criteria from entanglement-breaking channels with trace-preserving}
Observe the completely positive
linear map $\Pi$
\begin{eqnarray}
\sigma\mapsto \Pi[\sigma]=\sum_{n=1}^{\emph{d}_B}
|e_{n}\rangle\langle e_{n}|\sigma|e_{n}\rangle\langle e_{n}|,
\end{eqnarray}
where the set $\{|e_{n}\rangle\}$ form the orthonormal basis in
$\mathcal{H}_{B}$ and $\sum_{n}|e_{n}\rangle\langle e_{n}|=I$. Let
$P_{n}=|e_{n}\rangle\langle e_{n}|$, we can get the following linear
map
\begin{eqnarray}
\sigma\mapsto \Pi[\sigma]=\sum_{n=1}^{\emph{d}_{B}}P_{n}\sigma P_{n}
,
\end{eqnarray}
which is the mathematical description of the \emph{wave-packet
reduction}\cite{Benatti05}. It is not hard to see that Eq.(17) is
EBT.

 \textbf{Corollary 1.} A density operator $\rho$ is separable if its entries
$\rho_{ij,kl}(j\neq l)=0$, where $\rho_{ij,kl}=\langle ij|\rho
|kl\rangle$, $\{|i\rangle\}$ and $\{|k\rangle\}$ in
$\mathcal{H}_{A}$, $\{|j\rangle\}$ and $\{|l\rangle\}$ in
$\mathcal{H}_{B}$, are computational bases.

 \textbf{proof:} An arbitrary density operator $\rho
\in \mathcal{B}(\mathcal{H}_{A,B})^{+}$ can be defined as
\begin{eqnarray}
\rho=\sum_{ijkl}\langle ij|\rho |kl\rangle(|i\rangle\langle
k|\otimes |j\rangle\langle l|)
\end{eqnarray}
by computational (real orthonormal) bases $\{|i\rangle\}$ and
$\{|k\rangle\}$ in $\mathcal{H}_{A}$, $\{|j\rangle\}$ and
$\{|l\rangle\}$ in $\mathcal{H}_{B}$. By Eq.(17) and Eq.(19), we
have
\begin{eqnarray}
(I\otimes\Pi)[\rho]=\sum_{ijkl}\langle ij|\rho
|kl\rangle(|i\rangle\langle
k|\otimes(\sum_{n=1}^{\emph{d}_{B}}|e_{n}\rangle\langle
e_{n}|(|j\rangle\langle l|)|e_{n}\rangle\langle e_{n}|)).
\end{eqnarray}

Let the orthonormal basis $\{|e_{n}\rangle\}$ be the same as
$\{|l\rangle\}$, it follows that
\begin{eqnarray}
(I\otimes\Pi)[\rho]=\sum_{ikl}\langle il|\rho
|kl\rangle(|i\rangle\langle k|\otimes(|l\rangle\langle
l|))=\rho_{il,kl}.
\end{eqnarray}\hfill $\square$

This result establishes a class of the representations of matrix
form for the separable state output from the EBT.

According to corollary 1, for example, a two-qubit quantum state
\begin{eqnarray}
\rho=
    \left(
      \begin{array}{cccc}

                 \rho_{00,00}  &0 & \rho_{00,10} &0  \\

                 0 & \rho_{01,01} &0 & \rho_{01,11} \\

                 \rho_{10,00}  &0 & \rho_{10,10} &0 \\

                 0 & \rho_{11,01} &0 & \rho_{11,11}
      \end{array}
    \right)
\end{eqnarray}
is separable.

By a simple calculation, we can get different sufficient
separability criteria in matrix form with different
$\{|e_{n}\rangle\}$ in Eq.(20). For example, let
$|e_{0}\rangle=\frac{1}{2}(\sqrt{2}|0\rangle-|1\rangle+|2\rangle)$,
$|e_{1}\rangle=\frac{1}{2}(\sqrt{2}|0\rangle+|1\rangle-|2\rangle)$
and $|e_{2}\rangle=\frac{\sqrt2}{2}(|1\rangle+|2\rangle)$ for
qutrits, if
\begin{eqnarray}
\rho=&&\nonumber\\
 \left(
      \begin{array}{ccccccccc}

    \rho_{00,00} & \rho_{00,01} & \rho_{00,02} & \rho_{00,10} &\rho_{00,11}  & \rho_{00,12} & \rho_{00,20} & \rho_{00,21} & \rho_{00,22}\\

    \rho_{01,00} & \rho_{01,01} & \rho_{01,02} & \rho_{01,10} &\rho_{01,11}  & \rho_{01,12} & \rho_{01,20} & \rho_{01,21} & \rho_{01,22}\\

    \rho_{02,00} & \rho_{02,01} & \rho_{02,02} & \rho_{02,10} &\rho_{02,11}  & \rho_{02,02} & \rho_{02,20} & \rho_{02,21} & \rho_{02,22}\\

    \rho_{10,00} & \rho_{10,01} & \rho_{10,02} & \rho_{10,10} &\rho_{10,11}  & \rho_{10,12} & \rho_{10,20} & \rho_{10,21} & \rho_{10,22}\\

    \rho_{11,00} & \rho_{11,01} & \rho_{11,02} & \rho_{11,10} &\rho_{11,11}  & \rho_{11,12} & \rho_{11,20} & \rho_{11,21} & \rho_{11,22}\\

    \rho_{12,00} & \rho_{12,01} & \rho_{12,02} & \rho_{12,10} &\rho_{12,11}  & \rho_{12,02} & \rho_{12,20} & \rho_{12,21} & \rho_{12,22}\\

    \rho_{20,00} & \rho_{20,01} & \rho_{20,02} & \rho_{20,10} &\rho_{20,11}  & \rho_{20,12} & \rho_{20,20} & \rho_{20,21} & \rho_{20,22}\\

    \rho_{21,00} & \rho_{21,01} & \rho_{21,02} & \rho_{21,10} &\rho_{21,11}  & \rho_{21,12} & \rho_{21,20} & \rho_{21,21} & \rho_{21,22}\\

    \rho_{22,00} & \rho_{22,01} & \rho_{22,02} & \rho_{22,10} &\rho_{22,11}  & \rho_{22,02} & \rho_{22,20} & \rho_{22,21} & \rho_{22,22}\\

      \end{array}
    \right)
\end{eqnarray} is an arbitrary density operator (entangled or not), then

\begin{eqnarray}
\rho^{'}= \left(\begin{array}{c|c|c}

           \rho^{00} & \rho^{01} & \rho^{02}  \\

    \hline \rho^{10} & \rho^{11} & \rho^{12}  \\

    \hline \rho^{20} & \rho^{21} & \rho^{22}

\end{array}\right)
\end{eqnarray} is separable, where

\begin{eqnarray}
\rho^{i_Aj_A}=
 \left(\begin{array}{ccc}

    \rho^{i_Aj_A}_{00} & \rho^{i_Aj_A}_{01} & \rho^{i_Aj_A}_{02}  \\

    \rho^{i_Aj_A}_{10} & \rho^{i_Aj_A}_{11} & \rho^{i_Aj_A}_{12}  \\

    \rho^{i_Aj_A}_{20} & \rho^{i_Aj_A}_{21} & \rho^{i_Aj_A}_{22}

\end{array}\right)
\end{eqnarray}and

\begin{eqnarray}
\rho^{i_Aj_A}_{00}&=&
\frac{2\rho_{i_A0,j_A0}+\rho_{i_A1,j_A1}-\rho_{i_A1,j_A2}-
\rho_{i_A2,j_A1}+\rho_{i_A2,j_A2}}{4},\\
\rho^{i_Aj_A}_{01}&=&
\frac{\rho_{i_A0,j_A1}-\rho_{i_A0,j_A2}+\rho_{i_A1,j_A0}-\rho_{i_A2,j_A0}}{4},\\
\rho^{i_Aj_A}_{02}&=&
\frac{-\rho_{i_A0,j_A1}+\rho_{i_A0,j_A2}-\rho_{i_A1,j_A0}+\rho_{i_A2,j_A0}}{4},\\
\rho^{i_Aj_A}_{10}&=&
\frac{\rho_{i_A0,j_A1}-\rho_{i_A0,j_A2}+\rho_{i_A1,j_A0}-\rho_{i_A2,j_A0}}{4},\\
\rho^{i_Aj_A}_{11}&=&
\frac{2\rho_{i_A0,j_A0}+3\rho_{i_A1,j_A1}+\rho_{i_A1,j_A2}+\rho_{i_A2,j_A1}+3\rho_{i_A2,j_A2}}{8},\\
\rho^{i_Aj_A}_{12}&=&
\frac{-2\rho_{i_A0,j_A0}+\rho_{i_A1,j_A1}+3\rho_{i_A1,j_A2}+3\rho_{i_A2,j_A1}}{8},\\
\rho^{i_Aj_A}_{20}&=&
\frac{-\rho_{i_A0,j_A1}+\rho_{i_A0,j_A2}-\rho_{i_A1,j_A0}+\rho_{i_A2,j_A0}}{4},\\
\rho^{i_Aj_A}_{21}&=&
\frac{-2\rho_{i_A0,j_A0}+\rho_{i_A1,j_A1}+3\rho_{i_A1,j_A2}+3\rho_{i_A2,j_A1}}{8},\\
\rho^{i_Aj_A}_{22}&=&
\frac{2\rho_{i_A0,j_A0}+3\rho_{i_A1,j_A1}+\rho_{i_A1,j_A2}+\rho_{i_A2,j_A1}+3\rho_{i_A2,j_A2}}{8}
\end{eqnarray}for all $0\leq i_A,j_A\leq 2$.

Clearly, we can get different sufficient separability criteria in
\emph{matrix form} from different EBTs.

\subsection{The separability criteria from entanglement-breaking channels without trace-preserving}

Observe the depolarizing channel\cite{Nielsen00}
\begin{eqnarray}
D_\epsilon(\rho) &=& (1-\epsilon) \frac{I}{d_{A,B}} + \epsilon \rho.
\end{eqnarray}

Concretely, let us consider first the case of states of two qubits.
According to Ref. \cite{Braunstein99}, an arbitrary density operator
$\rho$ for two qubits can be written as
\begin{eqnarray}
\rho &=& {1\over 4}\Big[ ( \omega_i\omega_j +c_{i0}\omega_j +
\omega_ic_{j0} + c_{ij})
P_i\otimes P_j\nonumber\\
& &\quad + \big( \omega_i\omega_j -c_{i0}\omega_j + \omega_ic_{j0} -
c_{ij}\big)
\overline P_i\otimes P_j\nonumber\\
& &\quad + \big( \omega_i\omega_j +c_{i0}\omega_j - \omega_ic_{j0} -
c_{ij}\big)
 P_i\otimes \overline P_j\nonumber\\
& &\quad  + \big( \omega_i\omega_j - c_{i0}\omega_j - \omega_ic_{j0}
+ c_{ij}\big)
 \overline P_i\otimes \overline P_j \Big] \label{rho_representation}
\end{eqnarray}
and the maximally mixed density operator $I_4$ for two qubits may be
written as
\begin{eqnarray}
{1\over 4}\big( \omega_i\omega_j \big) \Big[
 P_i\otimes P_j
+ \overline P_i\otimes P_j +
 P_i\otimes \overline P_j
 +
 \overline P_i\otimes \overline P_j \Big],
\end{eqnarray}where $I_2$ is the $2\times 2$
identity matrix, $\omega_{i(j)}=1/3$, $-1\leqslant
c_{i(j)0}\leqslant 1$, $-1\leqslant c_{ij}\leqslant 1$,
\begin{eqnarray}
P_{i(j)} &=& {1\over 2} (I_2 + \sigma_{i(j)}),\nonumber\\
\overline P_{i(j)} &=& {1\over 2} (I_2 - \sigma_{i(j)}),\nonumber
\end{eqnarray}$\sigma_{i(j)}$ are the Pauli matrices and $i(j)=1,2,3$.

Let
\begin{eqnarray}
&&q_{0}^{ij}={1\over 4}( ( \omega_i\omega_j +\epsilon(c_{i0}\omega_j
+ \omega_ic_{j0} + c_{ij}))),\nonumber\\
&&q_{1}^{ij}={1\over 4}(( \omega_i\omega_j +\epsilon(-c_{i0}\omega_j
+ \omega_ic_{j0} -
c_{ij}))),\nonumber\\
&&q_{2}^{ij}={1\over 4}((
 \omega_i\omega_j +\epsilon(c_{i0}\omega_j -
\omega_ic_{j0} - c_{ij}))),\nonumber\\
&&q_{3}^{ij}={1\over 4}((
 \omega_i\omega_j+\epsilon(- c_{i0}\omega_j -
\omega_ic_{j0} + c_{ij}))).\nonumber
\end{eqnarray} We have
\begin{eqnarray}
D_\epsilon(\rho) &=& (1-\epsilon) \frac{I_4}{d_{A,B}} + \epsilon \rho\\
&=& q_{0}^{ij}
 P_i\otimes P_j
+ q_{1}^{ij}\overline P_i\otimes P_j +
 q_{2}^{ij}P_i\otimes \overline P_j
 +q_{3}^{ij}
 \overline P_i\otimes \overline P_j \\
&=&(I\otimes\Phi)((|00\rangle+|11\rangle)(\langle 00|+\langle 11|)),
\end{eqnarray} where $\Phi(\sigma)= P_j
\textrm{Tr}(\sigma q_{0}^{ij}P_i)+ P_j \textrm{Tr}(\sigma
q_{1}^{ij}\overline P_i)
 + \overline P_j\textrm{Tr}(\sigma q_{2}^{ij} P_i)
 +\overline P_j Tr(\sigma q_{3}^{ij}\overline P_i)$.

Clearly, $\sum_n q_{n}^{ij}=1$. Since $\Phi(\cdot)$ is an EBC if
$q_{n}^{ij}\geqslant 0$, i.e. $\epsilon\leqslant {1\over 15}$,
$D_\epsilon(\rho)$ is separable. Thus, the above result coincides
with the result of Ref. \cite{Braunstein99}.

By Eq.(35), we have
\begin{eqnarray}
D_\epsilon(\rho) &=& (1-\epsilon) \frac{I}{d_{A,B}} + \epsilon \rho\\
&=&(I\otimes\Phi)(|I\rangle\langle I|)\\
&=&\sum_{i,j}|i_A\rangle\langle j_A|\otimes\Phi(|i_B\rangle\langle
j_B|)\\
&=&\sum_{i_A,j_A}|i_A\rangle\langle
j_A|\otimes\sum_{i_B,j_B}\Phi_{i_A,j_A}(|i_B\rangle\langle j_B|)
\end{eqnarray}where
\begin{equation} \label{eq:1}
\Phi_{i_A,j_A}(|i_B\rangle\langle j_B|)=\left \{ \begin{array}{ccc}
         &\sum_{i_Bj_B}(\epsilon\langle i_Ai_B|\rho |j_Aj_B\rangle|i_B\rangle\langle j_B|+\frac{1-\epsilon}{d_{A,B}}|i_B\rangle\langle j_B|),&  i=j \\
         &\sum_{i_Bj_B}(\epsilon\langle i_Ai_B|\rho |j_Aj_B\rangle|i_B\rangle\langle j_B|),&                                                  i\neq j.
                 \end{array}          \right.
                          \end{equation}

Since the depolarizing channel in Eq.(35) is entanglement breaking
if $\epsilon \leq \frac{1}{d+1}$\cite{Grudka10},
$\Phi_{i_A,j_A}(\cdot)$ is EBC for all $0\leq i_A,j_A\leq d_A-1$ if
$\epsilon \leq \frac{1}{d+1}$.

Furthermore, for an arbitrary density operator $\rho$ (separable or
not),
\begin{eqnarray}
\rho
&=&(I\otimes\Psi)(|I\rangle\langle I|)\\
&=&\sum_{i_A,j_A}|i_A\rangle\langle
j_A|\otimes\sum_{i_B,j_B}\Psi_{i_A,j_A}(|i_B\rangle\langle j_B|)
\end{eqnarray}where
\begin{equation}
\Psi_{i_A,j_A}(|i_B\rangle\langle j_B|)=\sum_{i_B,j_B}\langle
i_Ai_B|\rho |j_Aj_B\rangle|i_B\rangle\langle j_B|
\end{equation}is a map for all $0\leq i_A,j_A\leq d_A-1$. By Eq.(48), we have
\begin{eqnarray}
\Psi_{i_A,j_A}(\sigma) &=&\Psi_{i_A,j_A}(\sum_{i_Bj_B}\langle
i_B|\sigma|j_B\rangle|i_B\rangle\langle
j_B|)\\
&=&\sum_{i_Bj_B}\langle i_Ai_B|\rho |j_Aj_B\rangle\langle
i_B|\sigma|j_B\rangle|i_B\rangle\langle j_B|,
\end{eqnarray}where $\sigma$ is any density operator in $\mathbb{C}^{d_B}$ and its entries $\sigma_{i_Bj_B}=\langle
i_B|\sigma|j_B\rangle|i_B\rangle\langle j_B|$.

An arbitrary bipartite state $\rho$ in $\mathbb{C}^{d_A}\otimes
\mathbb{C}^{d_B}$ may be considered as an $d_A\times d_A$ matrix
with the entries being $d_B\times d_B$ matrices.

Concretely, let
\begin{eqnarray}
\rho=  \left(\begin{array}{c|c|c|c}

                 \rho^{00}  &\rho^{01} &\ldots &\rho^{0(d_A-1)} \\

                 \hline \rho^{10}  &\rho^{11} &\ldots &\rho^{1(d_A-1)} \\

                 \hline \vdots &\vdots &\ddots & \vdots  \\

                 \hline \rho^{(d_A-1)0}  &\rho^{(d_A-1)1} &\ldots &\rho^{(d_A-1)(d_A-1)}
\end{array} \right),\
\end{eqnarray}where
\begin{eqnarray}
\rho^{i_Aj_A}=     \left(
      \begin{array}{cccc}

                 \rho^{i_Aj_A}_{00}  &\rho^{i_Aj_A}_{01} &\ldots &\rho^{i_Aj_A}_{0(d_B-1)} \\

                 \rho^{i_Aj_A}_{10}  &\rho^{i_Aj_A}_{11} &\ldots &\rho^{i_Aj_A}_{1(d_B-1)} \\

                 \vdots &\vdots &\ddots & \vdots  \\

                 \rho^{i_Aj_A}_{(d_B-1)0}  &\rho^{i_Aj_A}_{(d_B-1)1} &\ldots &\rho^{i_Aj_A}_{(d_B-1)(d_B-1)}
      \end{array}
    \right)
\end{eqnarray}for all $0\leq i_A,j_A\leq d_A-1$. Therefore,
we can get the following theorem.

 \textbf{Theorem 3.} A density operator $\rho$ is separable
 if all blocks $\rho^{i_Aj_A} (0\leq i_A,j_A\leq d_A-1)$ of $\rho$ are positive.

\textbf{proof:} By Eq.(50), Eq.(51) and Eq.(52), we have
 \begin{eqnarray}
\Psi_{i_A,j_A}(\sigma)=&&\nonumber\\
\left(
      \begin{array}{cccc}

                 \rho^{i_Aj_A}_{00}\cdot \sigma_{00}  &\rho^{i_Aj_A}_{01}\cdot \sigma_{01} &\ldots &\rho^{i_Aj_A}_{0(d_B-1)}\cdot \sigma_{0(d_B-1)} \\

                 \rho^{i_Aj_A}_{10}\cdot \sigma_{10}  &\rho^{i_Aj_A}_{11}\cdot \sigma_{11} &\ldots &\rho^{i_Aj_A}_{1(d_B-1)}\cdot \sigma_{1(d_B-1)} \\

                \vdots &\vdots &\ddots & \vdots  \\

                 \rho^{i_Aj_A}_{(d_B-1)0}\cdot \sigma_{(d_B-1)0}  &\rho^{i_Aj_A}_{(d_B-1)1}\cdot \sigma_{(d_B-1)1} &\ldots &\rho^{i_Aj_A}_{(d_B-1)(d_B-1)}\cdot \sigma_{(d_B-1)(d_B-1)}
      \end{array}
    \right)
\\
=\rho^{i_Aj_A}\circ\sigma,&&
\end{eqnarray}where $\rho^{i_Aj_A}\circ\sigma$ denotes the Hadamard product of $\rho^{i_Aj_A}$ and $\sigma$. By Schur product theorem\cite{Horn90}, since $\sigma$ is a density
operator in $\mathbb{C}^{d_B}$ and positive,
$\Psi_{i_A,j_A}(\sigma)$ is positive if $\rho^{i_Aj_A}$ is positive.
Since all $\rho^{i_Aj_A}$ are positive, all $\Psi_{i_A,j_A}(\sigma)$
are positive. It is not hard to see that all
$\Psi_{i_A,j_A}(\sigma)$ can be written in the form of Eq. (2), and
therefore $\rho$ is separable. \hfill $\square$

Note that the identity map up to a common factor on subsystem does
not mean the identity map on composite system for channel without
trace-preserving. For example,
\begin{eqnarray}
\rho= \frac{1}{4}\left(\begin{array}{cccc}

   1 & 1 & 1 & 1 \\
   1 & 1 & 1 & 1 \\
   1 & 1 & 1 & 1 \\
   1 & 1 & 1 & 1 \\

\end{array}\right).
\end{eqnarray} is separable.

Clearly, the state in theorem 3 is a family of completely new
separable PPT states (states which are positive under partial
transposition)\cite{M.Horodecki96}. By Fejer's theorem\cite{Horn90},
we have

 \textbf{Corollary 2.} A density operator $\rho$ is separable if
\begin{eqnarray}
\sum_{i_B,j_B} \rho^{i_Aj_A}_{i_Bj_B}\cdot \sigma_{i_Bj_B}\geqslant
0
\end{eqnarray} for all $0\leq i_A,j_A\leq d_A-1$ and any density operator
$\sigma$ in $\mathbb{C}^{d_B}$.

\section{Conclusion}
In conclusion, we have demonstrated that the origin of an arbitrary
separable state in arbitrary composite quantum systems of arbitrary
dimensions can originate from a maximally entangled state by the
EBC. A class of separability criteria can be obtained from the EBC
and a family of completely new separable PPT states is given. The
separability criteria from EBC without trace-preserving are under
investigation.

 \ack \indent We would like to appreciate Zhihao Ma, Guang Ping
He, Julio de Vicente and Mary Beth Ruskai for helpful discussions
and suggestions. We would like to thank anonymous helpful comments
and suggestions to improve the original manuscript. We are also
grateful to Daowen Qiu and Guang Ping He for their excellent
lectures on quantum computation and quantum information. This work
is in part supported by the Key Project of
NSFC-Guangdong Funds (No. U0935002). \\

\end{document}